\documentclass[letterpaper, 10 pt, conference]{ieeeconf}  

\usepackage{cite}[short&compress]
\usepackage{todonotes} 
\usepackage{float}
\usepackage{booktabs}
\usepackage{multirow}
\usepackage{csquotes}
\usepackage{hyperref} 
\usepackage{graphicx}
\usepackage{dblfloatfix}  
\usepackage{booktabs}

\usepackage{multirow}
\usepackage[table,xcdraw]{xcolor}
\usepackage{colortbl}
\usepackage[table,xcdraw]{xcolor}

\IEEEoverridecommandlockouts                              

\title{\LARGE \bf
When to Say "Hi" - Learn to Open a Conversation with an in-the-wild Dataset
}

\author{Michael Schiffmann$^{1}$, Felix Struth$^{2}$, Sabina Jeschke$^{3}$ and Anja Richert$^{1}$
\thanks{*The authors acknowledge the financial support by the Federal Ministry of Education and Research of Germany in the framework FH-Kooperativ 2-2019 (project number 13FH504KX9)}
\thanks{$^{1}$Michael Schiffmann and Anja Richert are with the Cologne Cobots Lab, TH Köln - University of Applied Sciences, 50679 Cologne, Germany
        {\tt\small michael.schiffmann,anja.richert@th-koeln.de}}%
\thanks{$^{2}$Felix Struth is a master graduate with the Cologne Cobots Lab, TH Köln - University of Applied Sciences, 50679 Cologne, Germany {\tt\small felixstruth@web.de}}%
\thanks{$^{3}$Sabina Jeschke is with KI Park e.V. \& FAU - Friedrich-Alexander University of Erlangen-Nuremberg}
}

\begin{document}

\maketitle
\thispagestyle{empty}
\pagestyle{empty}

\begin{abstract}
The social capabilities of socially interactive agents (SIA) are a key to successful and smooth interactions between the user and the SIA. A successful start of the interaction is one of the essential factors for satisfying SIA interactions. For a service and information task in which the SIA helps with information, e.g. about the location, it is an important skill to master the opening of the conversation and to recognize which interlocutor opens the conversation and when. 
We are therefore investigating the extent to which the opening of the conversation can be trained using the user's body language as an input for machine learning to ensure smooth conversation starts for the interaction. In this paper we propose  the Interaction Initiation System (IIS) which we developed, trained and validated using an in-the-wild data set.
In a field test at the Deutsches Museum Bonn, a Furhat robot from Furhat Robotics was used as a service and information point. Over the period of use we collected the data of \textit{N} = 201 single user interactions for the training of the algorithms. We can show that the IIS, achieves a performance that allows the conclusion that this system is able to determine the greeting period and the opener of the interaction.

\end{abstract}

\section{INTRODUCTION}

In our opinion, appropriate manners are of great importance in connection with service and information tasks. In this human-robot interaction (HRI) use case, a public space scenario, visitors to a museum interact with a Furhat robot, which is utilized as an information point that provides relevant information about the museum. Since interacting with a humanoid robot is likely to be unfamiliar to most visitors, this is a new experience. A socially interactive agent that is used as a service and information point should, therefore, be able to start the conversation in an appropriate and natural manner.
In a comparable situation in human communication, it is intuitively clear who is starting the conversation since people recognize the intention of the person they want to talk to  based on gestures, facial expressions, and posture \cite{axelsson_modeling_2022}.
In HRI, conversation initiation plays a crucial role in shaping the course of interactions and influencing user trust. Xu and Howard \cite{Xu} highlight that a robot's first impression has a significant impact on user trust. Paetzel \cite{Paetzel} found that initial judgments of a robot's competence often persist. Erel  \cite{Erel} shows that the manner of initiation affects willingness to assist.
It is, therefore, of great importance for a socially interactive agent to master the opening sequence of an interaction. We are investigating the extent to which the timing and type of greeting can be learned with the help of machine learning to make the interaction more natural. The SIA must be able to decide whether to wait for a greeting, to listen, or to take the initiative.
To investigate this, we use a dataset from a field experiment in a museum to train machine learning models that compose the Interaction Initiation System. 
In the following, we will give a brief overview about the related work, the used robotic system, and the setting of the experiments to collect the dataset. We then introduce the methods used to give information about the experiment's procedure and then introduce the interaction initiation system. Finally, we present the training and validation results and end with a conclusion. 

\section{Related Work}
Several approaches have been explored to determine the optimal timing of conversation initiation. Hall \cite{hall1966hidden} analyzed spatial distance as an indicator of social relationships in human-to-human communication. Other studies on human-robot interaction, such as Gehle \cite{Gehle}, used the Wizard of Oz paradigm in a laboratory study to investigate control opening parameters. Shi \cite{shi_spatial_nodate} developed a state machine model that used body position and gaze to determine timing in a laboratory setting.
In addition to timing, the manner of initiation is critical. Verbal greetings \cite{Fischer}  and nonverbal cues such as waving the hand, eye contact, and gestures \cite{Kahn,Heenan,SidnerLee} have been shown to be effective. Fischer \cite{Fischer} showed that incremental volume adjustments can improve user attention. Models such as "pause \& restart" improve the willingness to interact by interrupting and restarting the greeting \cite{Pitsch}. Observing social norms and using visual context can also optimize conversation initiation \cite{Brink,Janssens}.
These studies show how important a well-thought-out conversation initiation is for successful HRI.
Furthermore, the previous approaches show that explicit signals such as gaze were used in each case. However, the studies also show that there is a lack of findings from in-the-wild user studies in which spontaneous and unplanned interactions in real use case environments are investigated and utilized.

\section{SYSTEM AND SETTING}

\begin{figure} 
    \vspace{2.5mm}    
    \includegraphics[width=1\linewidth]{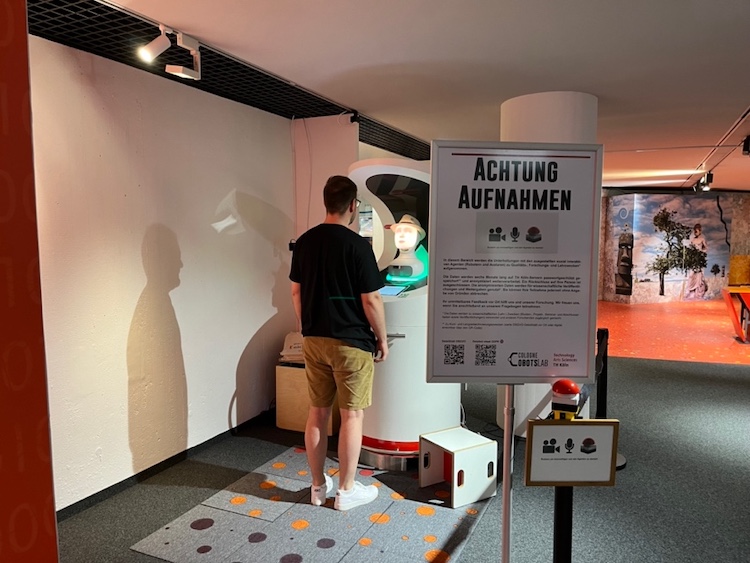}
    \caption{Experimental setup in the Deutsches Museum Bonn. View from the entrance area to the demarcated interaction area}
    \label{fig:System_Setting}
\end{figure} 

The experiment took place from July 23, 2024, to August 15, 2024, at the Deutsches Museum Bonn (DMB), during which data were collected. The DMB is a museum specializing in the presentation of artificial intelligence. The SIA robot used in this experiment was developed to answer visitors' questions about the exhibition. The robot system was placed near the entrance (see Fig. \ref{fig:System_Setting}). 
Visitors could interact with the robotic system within a designated interaction area. During interactions, video, and audio data recordings were made for research purposes, and the system persisted additional interaction data. Visitors must consent to data processing by pressing a red buzzer before entering the robot's interaction area. The robot remains in an idle state until consent is given, and it reminds visitors of the necessary data privacy agreement if consent has not been provided. If the user consented, the user gets detected when entering the interaction area and the robot is in a ready state to have a conversation.

The Furhat is mounted on a housing that includes a 10-inch tablet. During interactions, this tablet displays the transcribed dialogues. In addition to the camera integrated into the Furhat robot head, two additional wide-angle cameras are mounted on the sides. Labels behind the Furhat robot head explain the meaning of the LED lighting of the Furhat. The robot has been named Mira.

The system is divided into a front-end and a back-end. The front-end consists of the robot head running a so-called Furhat Skill, which uses the Furhat SDK for person management, speech-to-text, and text-to-speech. The connection to the back-end is facilitated via a REST API that enables the management of interaction sessions (start/end session, send utterances, maintenance commands). Person management is supported by one of the additional cameras, which uses YOLO (yolov5 \cite{jocherUltralyticsYolov5V312020})  person detection to verify the presence of individuals in the interaction area. Too tall or short individuals may not be consistently captured by the integrated camera, leading to early interaction termination.

The back-end is responsible for session and dialogue management and interaction data persistence. A unique session ID is generated for each interaction, to which all interaction data is assigned. Responses to user utterances are generated either via manually created, intent-based natural language understanding with a database or a fallback for intents beyond the crafted knowledge base, a locally executed LLM (Llama 3) generated answers.
The LLM receives the previous conversation and the prompt: "This is a conversation between a visitor of the german museum in bonn and a robot named mira. Mira is programmed to respond briefly and precisely, in no more than two sentences."
Additionally, the back-end controls video and audio recording, capturing footage from the two additional cameras and the camera integrated into the robot head.
 
\section{METHODOLOGY}
To provide a comprehensive overview, this section first outlines the experimental procedure, before transitioning to a detailed description of the Interaction Initiation System's approach.

\subsection{EXPERIMENT}
During the regular opening hours of the museum, visitors were able to interact with the robot. They were not given specific instructions on how to interact with the robot.  At the entrance to the separate area, they were informed about the experimental conditions, with a poster explaining the data protection modalities and pointing out the buzzer for consent. The visitors approached the robot independently and began the interaction. When entering the interaction area, the user is detected by the robot's person management and it is chosen randomly if the robot greets the user or if the robot starts in the listen mode to let the user start with the first utterance. This behavior supports the data acquisition for the IIS, to better understand the opening behavior.

\subsection{INTERACTION INITIATION SYSTEM}
Our approach to developing the Interaction Initiation System involved the following steps: First, the functional scope was defined. Subsequently we started with the feature engineering by viewing and annotating the recorded video data of the interactions and then the relevant features of the video frames got extracted. Finally, we trained time series regression and classification models to optimize the system.\\

\begin{figure*}[!ht]
    \includegraphics[width=\textwidth, keepaspectratio]{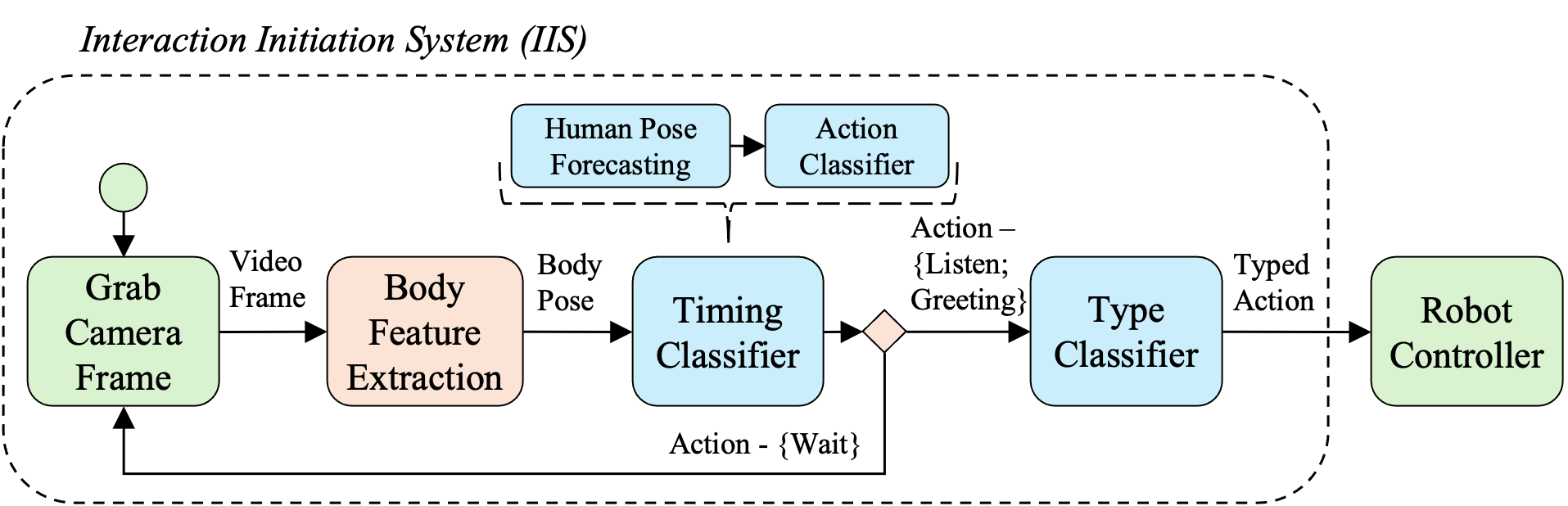}
    \caption{Flow Diagram of the Interaction Initiation System}
    \label{fig:IIS_System}
\end{figure*} 
\subsubsection{IIS - Function and Model}

The system is based on the separation of conversation initiation into two aspects: the time and the type. Figure \ref{fig:IIS_System} shows the program flow of the IIS. The blue boxes show the most relevant steps of IIS. The main components are the timing classifier and the type classifier. 
The IIS works primarily at the beginning of the interaction, when the user has consented to data protection and then approaches the robot. In this initial phase, the system registers the user and evaluates whether the robot greets the user, whether the user greets the robot first or whether it is better to wait to make a clear decision. If the user greets the robot first, a suitable greeting should be issued based on the user's utterance. It is therefore important to switch to listen mode in advance in order to be able to capture the user utterance correctly. For this reason the correct timing of the classification of greeting situation is an important factor for the IIS.
The focus of this paper is on determining the optimal time with the help of combined machine learning models. A model for determining the type has been developed theoretically, but has not yet been evaluated.
After the features have been extracted from the video data, the timing classifier of the system decides whether it is appropriate for the robot to open the conversation or whether the user will open the conversation promptly. This is realized by a two-stage workflow. A human pose forecasting model predicts the future position of the body points based on the extracted features. This predicted data is then used in the Action Classifier to classify a suitable action for the robot.
  
\begin{figure*}[!ht]
    \includegraphics[width=\textwidth]{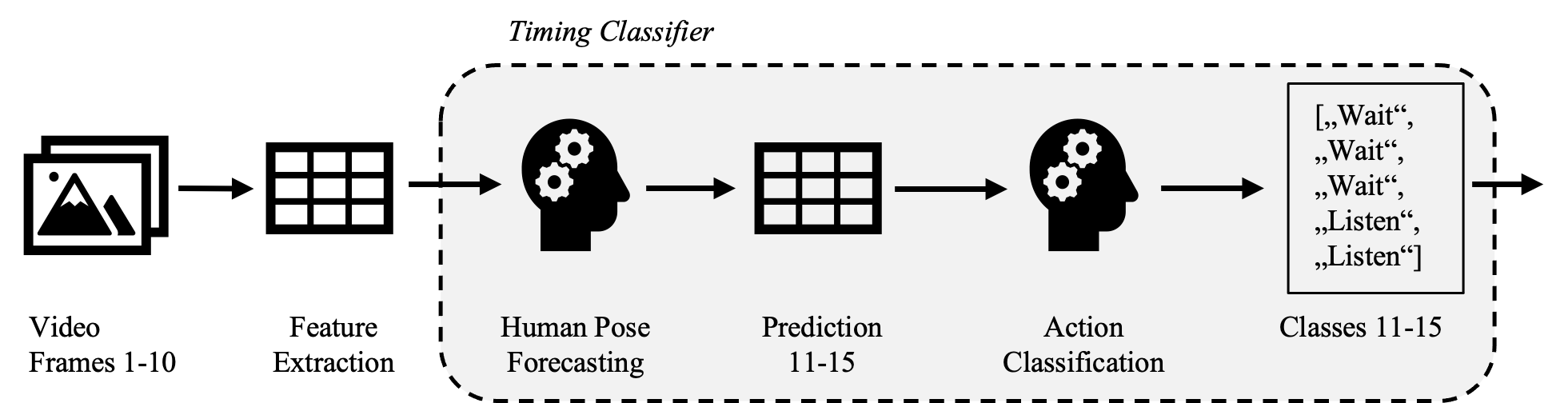}
    \caption{Detailed  View of the Timing Classifier}
    \label{fig:TimingClassifier}
\end{figure*}

The Human Pose Forecasting model is a BlockRNN time series model. It uses the last 10 instances of data to predict the next 5 (see Figure  \ref{fig:TimingClassifier} for the detailed sequence). 
In this setting, this allows a lead time of half a second based on the frame rate of 10 frames per second. The use of such a model is relevant for the recognition of user statements and the reaction time of the robot. As soon as the user opens their mouth to make a statement, the robot must already be in “listening” mode.

The Action Classifier is a support vector classifier trained with the features of the data set. The model assigns a data instance to one of the classes “Wait”, “Speak” or “Listen”. The decision is then discarded or passed on to the Type Classifier, depending on the class, as shown in Figure \ref{fig:IIS_System}. \\

\subsubsection{Feature Extraction}
Feature extraction is performed per frame and is essential to ensure the system's generalizability. Using only raw image data would only provide optimal results in specific, predefined contexts.

The extraction of the required features is used both for training the machine learning models and for the live system. For training, the recorded interactions are first pre-processed, and then the features are extracted as described below. The pre-processing of the video data includes the selection of suitable interactions based on defined guidelines and tailoring to the relevant part of the conversation, namely the approach of the user to the robot and the opening of the conversation in the form of an initial statement. The selection guidelines are:

\begin{itemize}
    \item Individual interactions as they exhibit more predictable user behavior compared to groups.
    \item The user must be fully visible in the front camera.
    \item The user is in the entrance area when the recording begins.
\end{itemize} 

To extract the features, Google MediaPipe \cite{mediaPipe}, a lightweight, hardware independent and fast framework to extract body postures from images, is used to identify a total of 543 key landmarks on the human body. These landmarks include 33 body points, 468 face points, and 21 points per hand. Each landmark is characterized by four parameters: the X, Y, and Z coordinates and the visibility. The X and Y coordinates are normalized values, while the Z coordinate represents the depth relative to the center of the hip. The visibility, scaled between 0 and 1, indicates the probability of recognizing a landmark on the image.

In addition to the key landmarks, a calculation of facial result objects is performed, which represent the facial expression shapes. Here, 52 different facial expression shapes are determined from the viewpoints, which are coded according to their probability values (0 to 1). The visibility of the points is primarily used as an indication of their presence. A threshold value 0.5 was set for the body landmarks, with landmarks with visibility values below this threshold being assigned zero coordinates. The visibility values of the face and hand points are not archived due to their constant zero values.

In order to use supervised machine learning models, each data instance must be given a label. For this purpose, each body feature instance of the interactions is annotated and assigned to one of the three classes “wait”, “speak” or “listen”. These classes describe which action is appropriate on the part of the robot. The annotation was performed by the authors and determined based on the review of the videos. For example, the class “Speak” is assigned if the user will not open the conversation in this interaction and the user is objectively close enough to the robot. The “Listen” class is assigned if the user will start the conversation and has approached the robot close enough. Although the user opens the conversation in 59\% of the interactions, there is more data that offers the robot the option of waiting or signals that the robot should open the conversation. The slight imbalance in the data is due to the naturally longer phases in which the user is too far away from the robot and very short and uncertain user greetings.

In summary, each data instance results in 1682 individual values, respectively, the features. This methodical recording and systematization of the data forms an essential basis for use in machine learning models to determine the optimal time to open a conversation.\\

\subsubsection{Training}

The pre-processing of the interactions resulted in a total number of 201 suitable interactions with 88 female and 113 male participants that were used to train and validate the proposed system. Since there is is no connection between the participants and an additional questionnaire no further demographic data is available.
The dataset consists of 26,675 labeled data instances with 1,682 features (see Table \ref{tab:tabDistri}). 

\begin{table}[!h]
\caption{Distribution of the dataset classes and distribution of train and test dataset. }
\centering
\begin{tabular}{@{}l|l|l|l|l@{}}
\cmidrule(lr){2-4}
                                     & Train Dataset  & Test Dataset & Total &  \\ \cmidrule(r){1-4}
\multicolumn{1}{|l|}{Instances}      & 23775 (89.1\%) & 2900 (10.9\%)  & 26675 &  \\ \cmidrule(r){1-4}
\multicolumn{1}{|l|}{Class "wait"}   & 11594 (89.9\%) & 1304 (10.1\%)  & 12898 &  \\ \cmidrule(r){1-4}
\multicolumn{1}{|l|}{Class "listen"} & 5829 (89.2\%)  & 713 (10.8\%)   & 6542  &  \\ \cmidrule(r){1-4}
\multicolumn{1}{|l|}{Class "speak"}  & 6352 (87.8\%) & 883 (12.2\%)  & 7235 &  \\ \cmidrule(r){1-4}
\end{tabular}
\label{tab:tabDistri}
\end{table}

The dataset is slightly unbalanced with respect to the “listen” class. This class imbalance is due to structural conditions within the use case and is difficult to avoid, given that there will always be longer time periods for the 'wait' and 'speak' classes. The split of the dataset results in a test dataset that comprises around 10.9\% of the data, which are 22 interactions 11 female and male each. The value ranges of the features from the body landmarks coordinates are between 0 and 1. Therefore, no further scaling of the data was performed.
The models for the timing classifier, the human pose forecasting model, and the action classifier were developed, trained, and initially evaluated using the train dataset. The human pose forecasting model is based on a dataset of 179 time series. The training process involved running 200 epochs to optimize model performance with each run containing 30 examples of a time series. Various base models and hyperparameters were tested during the training. The action classifier was trained using a list of different hyperparameter and different contributions of train and test data through GridSearchCV with 10 folds each. 

\section{RESULTS}

Our results consist of the Timing Classifier's test results and the components from which it is built: the Human Pose Forecasting Model and the Action Classifier. 

The Human Pose Forecasting model achieved a Root Mean Squared Error (RMSE) of 0.0426 for the BlockRNN base model, indicating high accuracy in predicting human poses after thorough training.

A support vector machine and a random forest classifier were compared with each other for the model selection of the action classifier. Both classifiers were optimized by GridSearchCV. This process identified the best hyperparameter listed in Table \ref{tab:hyperparameter}. As a result, the support vector machine classifier achieved an accuracy of 75.3\% on the test data while classifying the actions "wait," "speak," and "listen," utilizing all available features.

\begin{table}[!h]
\centering
\caption{Overview of the best Hyper-Parameter and used Features of the action classifier}
\begin{tabular}{@{}|l|l|l|l|@{}}
\toprule
Model                                                                                  & Hyper-Parameter                                                                                                                                          & Used Features                                                                       & Accuracy \\ \midrule
\multirow{2}{*}{\begin{tabular}[c]{@{}l@{}}Random\\ Forest \\ Classifier\end{tabular}} & \multirow{2}{*}{\begin{tabular}[c]{@{}l@{}}n\_estimators: 100, \\ max\_features: sqrt'', \\ bootstrap: True, \\ class\_weight: balanced''\end{tabular}} & \begin{tabular}[c]{@{}l@{}}All Features\\  $(\sum=1682)$\end{tabular} & 73.2\%   \\ \cmidrule(l){3-4} 
                                                                                       &                                                                                                                                                         & \begin{tabular}[c]{@{}l@{}}Only Facial \\  $(\sum=1404)$\end{tabular}   & 70.1\%   \\ \midrule
\multirow{2}{*}{\begin{tabular}[c]{@{}l@{}}Support \\ Vector \\ Machine\end{tabular}}  & \multirow{2}{*}{\begin{tabular}[c]{@{}l@{}}C: 1, gamma: scale'', \\ kernel: rbf'',\\  class\_weight: balanced''\end{tabular}}                           & \begin{tabular}[c]{@{}l@{}}All Features\\  $(\sum=1682)$\end{tabular} & 75.3\%   \\ \cmidrule(l){3-4} 
                                                                                       &                                                                                                                                                         & \begin{tabular}[c]{@{}l@{}}Only Facial \\ $(\sum=1404)$\end{tabular}   & 69.5\%   \\ \bottomrule
\end{tabular}
\label{tab:hyperparameter}
\end{table}

Since accuracy is less meaningful for unbalanced data sets, we consider weighted and macro-averages of precision, recall, and F1-Score. 
We validated the classification model twice: once separately from the Human Pose Forecast Model, and once together. Tables \ref{tab:TabActionClassConf} and \ref{tab:TabTimingClassConf} present both confusion matrices of the validations. In a slightly unbalanced test data set with 2,900 instances, the confusion matrix of the action classifier indicates that all classes were correctly classified in more than half of the cases, with the "wait" class leading the best and the “listen” class yielding the poorest results. The combination of both developed models in the timing classifier reflects a similar trend; however, the “listen” class is classified slightly worse, and the “speak” class is classified slightly better.

\begin{table}[!h]
\centering
\caption{Resulting 3x3 Confusion Matrix of the Action Classifier (SVC)}
\begin{tabular}{@{}ll|lll|@{}}
\cmidrule(l){3-5}
                                                        &          & \multicolumn{3}{l|}{Correct Class}                                                                                                                     \\ \cmidrule(l){3-5} 
                                                        &          & \multicolumn{1}{l|}{"listen"}                                         & \multicolumn{1}{l|}{"speak"}                     & "wait"                      \\ \midrule
\multicolumn{1}{|l|}{}                                  & "listen" & \multicolumn{1}{l|}{\cellcolor[HTML]{32CB00}{\color[HTML]{000000} 396}} & \multicolumn{1}{l|}{\cellcolor[HTML]{FE0000}285}  & \cellcolor[HTML]{FE0000}64   \\ \cmidrule(l){2-5} 
\multicolumn{1}{|l|}{}                                  & "speak"  & \multicolumn{1}{l|}{\cellcolor[HTML]{FE0000}302}                       & \multicolumn{1}{l|}{\cellcolor[HTML]{32CB00}575} & \cellcolor[HTML]{FE0000}26  \\ \cmidrule(l){2-5} 
\multicolumn{1}{|l|}{\multirow{-3}{*}{Predicted Class}} & "wait"   & \multicolumn{1}{l|}{\cellcolor[HTML]{FE0000}15}                        & \multicolumn{1}{l|}{\cellcolor[HTML]{FE0000}23}   & \cellcolor[HTML]{32CB00}1214 \\ \bottomrule
\end{tabular}
\label{tab:TabActionClassConf}
\end{table}

\begin{table}[!h]
\centering
\caption{Resulting 3x3 Confusion Matrix of the Timing Classifier which combines the Human Pose Forecasting and the Action Classifier}
\begin{tabular}{@{}ll|lll|@{}}
\cmidrule(l){3-5}
                                                        &          & \multicolumn{3}{l|}{Correct Class}                                                                                                                     \\ \cmidrule(l){3-5} 
                                                        &          & \multicolumn{1}{l|}{"listen"}                                         & \multicolumn{1}{l|}{"speak"}                     & "wait"                      \\ \midrule
\multicolumn{1}{|l|}{}                                  & "listen" & \multicolumn{1}{l|}{\cellcolor[HTML]{32CB00}{\color[HTML]{000000} 318}} & \multicolumn{1}{l|}{\cellcolor[HTML]{FE0000}234}  & \cellcolor[HTML]{FE0000}80   \\ \cmidrule(l){2-5} 
\multicolumn{1}{|l|}{}                                  & "speak"  & \multicolumn{1}{l|}{\cellcolor[HTML]{FE0000}380}                       & \multicolumn{1}{l|}{\cellcolor[HTML]{32CB00}634} & \cellcolor[HTML]{FE0000}42  \\ \cmidrule(l){2-5} 
\multicolumn{1}{|l|}{\multirow{-3}{*}{Predicted Class}} & "wait"   & \multicolumn{1}{l|}{\cellcolor[HTML]{FE0000}15}                        & \multicolumn{1}{l|}{\cellcolor[HTML]{FE0000}15}   & \cellcolor[HTML]{32CB00}1182 \\ \bottomrule
\end{tabular}
\label{tab:TabTimingClassConf}
\end{table}

\begin{table}[!h]
\centering
\caption{Resulting Metrics of the Timing Classifier. Table is showing the weighted and Macro Averages for Precision, Recall and the resulting F1-Score.}
\begin{tabular}{@{}llll|l|@{}}
\cmidrule(l){2-5}
\multicolumn{1}{l|}{}                  & \multicolumn{1}{l|}{Precision} & \multicolumn{1}{l|}{Recall} & F1-Score & Accuracy \\ \midrule
\multicolumn{1}{|l|}{Macro Average}    & \multicolumn{1}{l|}{69\%}    & \multicolumn{1}{l|}{69\%}   & 69\%   &          \\ \midrule
\multicolumn{1}{|l|}{Weighted Average} & \multicolumn{1}{l|}{74\%}    & \multicolumn{1}{l|}{74\%} & 74\%   &          \\ \midrule
                                       &                                &                             &          & 73.6\%     \\ \cmidrule(l){5-5} 
\end{tabular}
\label{tab:MetricsTable}
\end{table}

The main component of the IIS, the Timing Classifier is evaluated for further analysis on the test data of 22 already extracted recordings. As shown in Figure \ref{fig:TimingClassifier}, the timing classifier requires the extracted features of the recordings. These features are then processed in the Human Pose Forecasting model and the predicted output is classified using the Action Classifier.
To ensure the evaluation reflects reality as closely as possible the Timing Classifier is tested like in real deployment.
This means that the Human Pose Forecasting Model requires 10 data instances to start forecasting. Here, we classify the first 10 data instances directly with the action classifier and then switch to using both models consecutively as the forecast model starts to predict poses. This means the timing classifier is run through completely from the 11th data instance. During each loop, the final predicted class is recorded and compared against the actual classes.
The resulting Scores are shown in Table \ref{tab:MetricsTable}. The weighted precision stands at 74\%, with recall also at 74\%, leading to an F1 score of 74\%. However, the macro averages show that one class is predicted less accurately, as indicated by the F1 score of 69\%, suggesting a partially unstable classification. As can be seen from the confusion matrices, this appears to be the "listen" class, which requires further development. The lower prediction quality appears to be due to the minimal differences between the "listen" class and the others. The "listen" class is distinguished only by minimal mouth movements, which are difficult to recognize and predict. 
For the entire Interaction Initiation System, which includes the Type Classifier, further evaluation can be pursued through subsequent field tests, presenting valuable opportunities for continued assessment and refinement.

\section{CONCLUSIONS AND FUTURE WORK}
In this paper, we introduced the Interaction Initiation System, which was developed to predict the appropriate timing and type of greetings to make interactions with SIA in public spaces more natural. Human-robot interaction data were collected in a field study in-the-wild and used to train the main component, the Timing Classifier.

We demonstrated that the IIS enables a prediction of conversation-opening timing through user body posture classification based on an in-the-wild dataset.  This is achieved by first performing pose forecasting and then classifying the action ("listen", "wait", "speak"). The results for the unweighted F1-score in the current dataset is 69\%. We assess that the current approach would likely yield better performance with a larger dataset, as there would be sufficient training examples for all classes. 
We found that the pose recognition is not always robust resulting in noisy pose values, which aligns with the findings of \cite{Dill2023_MP} that the recording angle and movement type affects the detection accuracy. 
We assume that more stable recognition would benefit the system. The generation of training data in the field provides the significant advantage of using users' unbiased behavior for training. However, it also has a disadvantage: obtaining a balanced data set with a balanced class distribution depends on the number of trial days, as there is no control over when a user with a particular greeting preference interacts with the SIA. An additional issue is that user preferences, i.e., whether they want to be proactive greeters or be greeted by the robot, are not necessarily uniformly distributed in society. One limitation of this study is that we did not evaluate the "Type Classifier" component for selecting the greeting type. Future studies should therefore investigate whether the greeting behavior is perceived as natural and whether the selected greeting formula is perceived as appropriate.

In terms of transferability to other HRI scenarios, the approach is conceptually suitable for scenarios with a stationary robot. The data protection regulation requires the prior consent of the user, which, depending on the installation location and design of the robot's interaction area, leads to a change in the process of opening the conversation. For example, it could be that the user is already standing directly in front of the robot, gives his consent and starts the interaction directly. This would require a new data set in order to obtain the individual user behavior as training data.
In scenarios with a mobile robot, the approach is transferable to a limited extent, as a changing camera perspective influences the recording of the user and therefore larger amounts of data are required.

Approaches for further improving the Interaction Initiation System can be systematically found along the machine learning tool-chain, e.g., by explicit use of user-to-robot distance values or by employing a video dataset with significantly higher frames per second, which would allow better recognition of the user and especially of user speech movements. 
Additionally, it might be beneficial to choose an approach that uses user simulation with synthetic data based on real training data to achieve better results.

\addtolength{\textheight}{-12cm}   



\section*{ACKNOWLEDGMENT} 
The Research and Development activities were reviewed and approved by the Ethics
Research Committee of TH Köln (application no. THK-2023-
0004). The authors acknowledge the financial support by the
Federal Ministry of Education and Research of Germany in the
framework FH-Kooperativ 2-2019 (project number
13FH504KX9). We thank our collaboration partner DB Systel GmbH, the
Deutsches Museum Bonn, for their assistance and
contributions.

\bibliographystyle{IEEEtran}
\bibliography{Literatur}

\end{document}